\documentclass[aps,pra,onecolumn,10pt]{revtex4-1}
\usepackage[dvipdfm]{graphicx}
\usepackage[usenames,dvipsnames]{xcolor}
\usepackage[dvipdfm]{hyperref}
\usepackage{amsmath}
\begin{document}

\title{A Potential Based Quantization Procedure of the Damped Oscillator}
\author{Ferenc M\'arkus}
	\email[Corresponding author: ]{markus.ferenc@ttk.bme.hu; markus@phy.bme.hu}
	\affiliation{Department of Physics,
		Budapest University of Technology and Economics,
		H-1111 Budafoki \'ut 8., Budapest, Hungary}

	\author{Katalin Gamb\'ar}
	\email[ ]{gambar.katalin@uni-obuda.hu}
	\affiliation{Department of Natural Sciences, Institute of Electrophysics, \\ Kálmán Kandó Faculty of Electrical Engineering, \\ Óbuda University, \\ Tavaszmező u. 17, H-1084 Budapest, Hungary}
    
    \affiliation{Department of Natural Sciences, \\ National University of Public Service, \\  Ludovika t\'er 2, H-1083 Budapest, Hungary}

\date{\today}

\begin{abstract}
Nowadays, two of the most prospering fields of physics are quantum computing and spintronics. In both, the loss of information and dissipation plays a crucial role. In the present work we formulate the quantization of the dissipative oscillator, which aids understanding of the above mentioned, and creates a theoretical frame to overcome these issues in the future. Based on the Lagrangian framework of the damped spring system, the canonically conjugated pairs and the Hamiltonian of the system are obtained, by which the quantization procedure can be started and consistently applied. As a result, the damping quantum wave equation of the dissipative oscillator is deduced, by which an exact damping wave solution of this equation is obtained. Consequently, we arrive at such an irreversible quantum theory by which the quantum losses can be described.
\end{abstract}

\maketitle

\section{Introduction}

The existence of oscillator motion and the wave propagation mode is the necessary condition in the signal transmission, consequently, in the information transfer. In a realistic quantum operation, the dissipation of a signal appears due to loss of energy from a quantum system, like a single atom coupled to a single mode of electromagnetic radiation undergoes spontaneous emission \cite{nielsen2010}. This may happen in an ion trap quantum compution was proposed by Cirac and Zoller \cite{zoller95}. Similarly, in spin-wave interconnets, in spin-wave memories or spin-wave transducers the attenuation reduces the efficiency, the lossy spin-wave propagation leads to fundamental limitations \cite{mahmoud2020}. In the construction of quantum computers and further quantum information systems the qubits are responsible for information transfer. A single electron on solid neon surface gives the experimental realization of a new qubit platform \cite{zhou2022}. In this realization the limited coherent time is due to the energy and phase loss originated from the material surface deficiency and the noise of environment. 

These experiments and their theoretical discussions suggest the deeper understanding of dissipation of single quantum package. The key for the wave propagation is the oscillator, thus we need to find an uncontroversial description of the quantized damped oscillator.

Generally, the lack of success of the solution was due to the several stumbling blocks in the construction. The first difficulty was immediately in the formulation of Lagrangian, consequently, it is not possible to deduce the canonical variables, the Hamiltonian and the Poission bracket expressions. Thus the commutation rules, required for the quantization procedure, could not be formulated at all. \\  
The problem of the missing Lagrangian structure of the dissipative systems was much older than the elaboration of the quantization procedure of conservative systems, this goes back in time to Rayleigh \cite{rayleigh1877theory}. The equation of motion (EOM) of the harmonic oscillator
\begin{equation}
\ddot{x} + \omega^2 x = 0
\end{equation}
is deduced from the Lagrangian of 
\begin{equation}
L = \frac{1}{2} \dot{x}^2 - \frac{1}{2} \omega^2 x^2 ,
\end{equation}
where the related Euler-Lagrange equation is
\begin{equation}
\frac{{\mathrm{d}}}{{\mathrm{d}} t} \frac{{\mathrm{d}} L}{{\mathrm{d}} \dot{x}} - \frac{{\mathrm{d}} L}{{\mathrm{d}} x} = 0.
\end{equation}
To describe the damped oscillator, Rayleigh introduced the so-called dissipation potential --- pertaining to the drag force $F = -C \dot{x}$ --- 
\begin{equation}
\Phi = \frac{1}{2} C \dot{x}^2,
\end{equation}
by which the EOM could be recovered in the following way:
\begin{equation}  
\frac{{\mathrm{d}}}{{\mathrm{d}} t} \frac{{\mathrm{d}} L}{{\mathrm{d}} \dot{x}} - \frac{{\mathrm{d}} L}{{\mathrm{d}} x} = \frac{{\mathrm{d}} \Phi}{{\mathrm{d}} \dot{x}}.  \label{rayleigh_potential}
\end{equation}
It is easy to check that the correct EOM appears:
\begin{equation}
\ddot{x} + C \dot{x} + \omega^2 x = 0.
\end{equation}
However, it can be proven that the added term on the right side in Eq. (\ref{rayleigh_potential}) is not from the least action principle, and the related variational calculus, i.e., the Lagrangian frame is lost. Much later Bateman \cite{bateman1931} suggested a mirror image description in which a complementary equation appears due to the introduced function $y$. Here, the variation problem is
\begin{equation}
\delta \int y \left( \ddot{x} + C \dot{x} + \omega^2 x \right) dt = 0.
\end{equation}   
In addition to the damped oscillator equation, the mirror image equation is
\begin{equation}
\ddot{y} - C \dot{y} + \omega^2 y = 0.
\end{equation}
While the equation of $x$ relates to the damping solution, the equation of $y$ pertains to an exponentially increasing amplitude motion. The calculated Hamiltonian includes both of these functions at the same time. It may cause cumbersome explanation and elaboration of canonical variable pairs. A few years ago, Bagarello \emph{et al}. proved that the canonical quantization for the damped harmonic oscillator using the Bateman Lagrangian does not work \cite{bagarello2019}. Similarly, Morse and Feshbach \cite{morse1953} used the variable duplication method for the diffusion problem. In this case, the diffusion variable is considered a complex quantity, and its complex conjugate pair is the duplicated variable. Here, interpretation is cumbersome, due to the complex diffusion functions and the related canonical formulation. 

The idea of quantizing the damped oscillator, and the description of non-conserving energy subsystems coincide with the born of quantum theory itself \cite{caldirola1941,kanai1948}. Based on their original idea about quantum dissipation, the description has been strongly developed \cite{choi2013}. In other research, the explicit time-dependent formulation can also give a successful deduction of the dissipative oscillator \cite{dekker1981,dittrich1996}. However, the uncertainty principle is incompatible with the time-dependent mathematical structure \cite{weiss2012}. 

The dissipative quantum systems can be modeled by the sets of decoupled harmonic oscillators in a reservoir \cite{leggett1984,caldeira1993,rosenau2000}. The considered systems have a statistical behavior, so the observed dynamics differ from the motion of a single damped quantum oscillator. Understanding the quantum behavior of different dissipative processes is essential for modern research today \cite{markus2005,markus2021}.
 
The application of the complex absorbing potentials --- firstly used for the description of scattering processes \cite{Razavy2005,Vibok1991-2,Vibok1992,Halasz2000,Vibok2001,Halasz2003,Muga2004,Henderson2006,markus2016} --- may give a good chance in the descrition of dissipation and ireversibility in the quantum theory. The method is that the Schr\"odinger equation is formulated by a complex potential $V_r(x) + \mathrm{i} V_{c}(x)$, where $V_r(x)$ represents the conservative potential, and $V_{c}(x)$ pertains to the damping. Therefore, the EOM for this system is
\begin{equation}  \label{complex_schr_eq}
\frac{\hbar}{\mathrm{i}} \frac{ \partial \Psi }{\partial t} - \frac{\hbar^2}{2m} \nabla^2 \Psi + \left(V_{r} + \mathrm{i} V_{c}\right) \Psi = 0.
\end{equation} 
The deduced balance equation for $\Psi^{\ast} \Psi$ is
\begin{equation}  \label{complex_schr_rho}
\frac{\partial \left(\Psi^{\ast} \Psi \right)}{\partial t} + \frac{\mathrm{i} \hbar}{2m} \nabla \left(\Psi \nabla \Psi^{\ast} - \Psi^{\ast} \nabla \Psi\right)
- \frac{2}{\hbar}V_{c} \Psi^{\ast}\Psi = 0,
\end{equation}
where just the complex part of the potential remains, generating the loss of the system. An obvious choice is to describe the motion of the quantum damped oscillator by the complex harmonic potential introduced by real-valued angular frequencies $\omega_r$ and $\omega_c$ \cite{markus2016}
\begin{equation}  \label{nagative_imaginary_potential}
V(x) = \frac{1}{2} m \left( {\omega_r^2} - \mathrm{i} {\omega_c^2} \right) x^2.
\end{equation}
The solution can be obtained by the application of the Feynman path integral method \cite{feyn1,feyn2,khan,Dittrich2001}, as it was shown previously \cite{markus2016}. We see that this method stands on the complex generalization of the acting potential, and the dissipation appears as a consequence of this non-Hermitian potential. However, we miss the direct --- introduced by an "equation-level" --- formulation of the dissipation. 
  
An explicit time-dependent Lagrangian method using the WKB approximation in the quantization procedure was developed by Serhan \emph{et al}. \cite{serhan2018}. Despite the exponentially decreasing time-dependence of the wave function, it describes a standing solution in space. 

A path integral method with a dynamical friction term is suggested for quantum dissipative systems by El-Nabulsi \cite{elnabulsi2020}. Here, Stokes’ drag force introduces the loss. However, the equation does not contain the velocity-dependent term. It contrasts with the standard EOM, which yields the usual exponential relaxation in time. 
 
Presently, we apply the canonical quantization method for the damped harmonic oscillator. We point out this classically developed procedure works in dissipative cases, not only by conservative potentials. We start from the EOM, and we formulate the Lagrangian and the Hamiltonian of the problem in general in Sec. \ref{Sec_Lagrangian}. As a particular case, the Hamiltonian of the underdamped oscillator is expressed in Sec. \ref{Sec_H=0}. The canonical quantization procedure can be followed in Sec. \ref{quantization}, the damping wave function is calculated by the path integral method in Sec. \ref{path_integral}. The results are summarized in a short conclusion in Sec. \ref{conclusion}. \\ 
The present technique has multiple advantages compared to the previous approaches. i) The canonical expressions and the quantization steps are familiar with the usual procedure. Thus the developed description can be considered a generalization. ii) The solving methods, like the path integral method, can be applied without radical changes. iii) The required necessary difference in the interpretation of the damped wave function can be interpreted.

\section{Lagrangian and Hamiltonian of a damped harmonic oscillator}  \label{Sec_Lagrangian}

The quantization procedure requires the formulation of the complete Lagrangian-Hamiltonian frame first. To achieve this aim, we start our examination from the EOM for the damped harmonic oscillator
\begin{equation}
\ddot x + 2\lambda \dot x + \omega^2 x = 0,
\end{equation}
where $m$ is the mass, $\lambda$ is a specific damping factor, and $\omega$ is the angular frequency. By the measurable quantity $x$ we define a generator potential $q$, i.e., the definition equation can be obtained as \cite{szegleti2020,gambar1994,gambar2020}, 
\begin{equation}
x = \ddot q - 2\lambda \dot q + \omega^2 q.  \label{potencial_definition}
\end{equation}
A suitable Lagrangian can be formulated by the potential 
\begin{equation}
L = \frac{1}{2} \left( \ddot q - 2\lambda \dot q + \omega^2 q \right)^2.
\end{equation}
The equations of motion can be calculated from a Lagrangian of the general form  
\[
L\left(q, \dot q, \ddot q, \dots , q^{(n)}, \dots , q^{(N)}, t \right) 
\] 
($q^{(n)}$ denotes the $n$th order time derivative) \cite{courant1966}:
\begin{equation}
0 = \sum\limits_{n=0}^N (-1)^n \frac{{\mathrm{d}}^n}{{\mathrm{d}}t^n} \frac{\partial L}{\partial q^{(n)}}.
\end{equation}
This method results in the EOM of the harmonic oscillator for the potential as  Euler--Lagrange equation. In general, Hamiltonian formalism requires canonical coordinate and momentum pairs
\begin{eqnarray}
q_{n}&=&q^{(n-1)} \\
p_{n}&=&\sum\limits_{k=0}^{N-n}(-1)^k \frac{{\mathrm{d}}^k}{{\mathrm{d}}t^k} \frac{\partial L}{\partial q^{(n+k)}},
\end{eqnarray}
\noindent where $\left(n = 1, \dots, N \right)$. The Hamiltonian can be deduced from the above-mentioned general Lagrangian as
\begin{equation}
H = \sum\limits_{n=0}^{N} p_{n} \frac{{\mathrm{d}}q_n}{{\mathrm{d}}t} - L  \label{calculate_hamiltonian}
\end{equation}	
For the present particular case, $N=2$ and $n=1,2$, we obtain the relevant coordinates as
\begin{equation}
q_1 := q ,
\end{equation}
and
\begin{equation}
q_2 := \dot{q} .
\end{equation}
Moreover, the general expression for the momentum $p_1$ is
\begin{equation}
p_1 := \sum\limits_{k=0}^{1}(-1)^k \frac{{\mathrm{d}}^k}{{\mathrm{d}}t^k} \frac{\partial L}{\partial q^{(1+k)}} =
\frac{\partial L}{\partial \dot{q}} - \frac{{\mathrm{d}}}{{\mathrm{d}}t} \frac{\partial L}{\partial \ddot{q}},
\end{equation}
by which we calculate the particular case as
\begin{equation}
p_1 := 4 \lambda^2 \dot{q} - \dddot{q\hspace{0pt}} - \omega^2 \dot{q} -2 \lambda \omega^2 q .
\end{equation}
Similarly, we formulate the momentum $p_2$
\begin{equation}
p_2 := \sum\limits_{k=0}^{0}(-1)^k \frac{{\mathrm{d}}^k}{{\mathrm{d}}t^k} \frac{\partial L}{\partial q^{(2+k)}} =
\frac{\partial L}{\partial \ddot{q}},
\end{equation}
i.e.,
\begin{equation}
p_2 := \ddot{q} -2 \lambda \dot{q} + \omega^2 q.
\end{equation}
The Hamiltonian can be calculated by Eq. (\ref{calculate_hamiltonian}). As a first step, we obtain it by the potential function $q$ 
\begin{align}
H = p_1 \dot{q}_1 + p_2 \dot{q}_2 - L  \nonumber \\ = 2 \lambda^2 \dot{q}^2 - \dddot{q\hspace{0pt}} \dot{q} - \omega^2 \dot{q}^2 + \frac{1}{2} \ddot{q}^2 - \frac{1}{2} \omega^4 q^2.  \label{oscillator_hamiltonian}
\end{align}
The following step is substituting the potential function with the coordinates and the momenta. We arrive at the canonical formulation of the Hamiltonian as
\begin{equation}
H = \frac{1}{2} p_2^2 - \omega^2 p_2 q_1 + p_1 q_2 + 2 \lambda p_2 q_2.  \label{oscillator_hamiltonian_by_momenta}
\end{equation}
To preserve the energy-like unit of the Hamiltonian, we transform the coordinates and the momenta. The transformation means a simple product by $\omega$, $m \omega$ and $m \omega^2$. Thus, we obtain new coordinates $Q_1$ and $Q_2$, and new momenta $P_1$ and $P_2$. 
\begin{eqnarray}
p_2 &\longrightarrow& P_2 = m \omega p_2, \,\,\,\,\, [P_2] = \text{kg} \frac{\text{m}}{\text{s}},  \label{P_2}  \\
q_2 &\longrightarrow& Q_2 = \omega q_2, \,\,\,\,\, [Q_2] = \text{m}  \label{Q_2},  \\
p_1 &\longrightarrow& P_1 = m \omega p_1, \,\,\,\,\, [P_1] = \text{kg} \frac{\text{m}}{\text{s}^2},  \label{P_1}  \\
q_1 &\longrightarrow& Q_1 = \omega q_1, \,\,\,\,\, [Q_1] = \text{m}\,\text{s},  \label{Q_1}
\end{eqnarray}
Moreover, the Hamiltonian, $H'$, is obtained 
\begin{equation}
H'  \longrightarrow H' = m \omega^2 H, \,\,\,\,\, [H] = \text{kg} \frac{\text{m}^2}{\text{s}^2} = \text{J}.  \label{H'}
\end{equation}
The units of the quantities are denoted by the bracket $[ \quad ]$. Finally, the Hamiltonian of the damped oscillator is
\begin{equation}
H' = \frac{1}{2m} P_2^2 - \omega^2 P_2 Q_1 + P_1 Q_2 + 2 \lambda P_2 Q_2.  \label{oscillator_hamiltonian_by_new_momenta}
\end{equation}
Before we turn towards the quantization procedure, it is worth examining a further property of this Hamiltonian.

\section{The devil is in the details}  \label{Sec_H=0}

Since the formulation of Lagrangian does not contain explicit time dependence, thus the Hamiltonian must be a constant value, i.e., the Hamiltonian expresses a conservation law. At this point, it is an open question what the Hamiltonian exactly means. Here, we try to clarify the role of the Hamiltonian in this theory. \\
We focus just on the solutions that pertain to the underdamped and overdamped cases. As was shown previously by Szegleti \emph{et al}. \cite{szegleti2020}, the $q(t)$ solution holds
\begin{equation}
q(t) = a_1\mathrm{e}^{-(\lambda+\gamma)t} + a_2\mathrm{e}^{-(\lambda-\gamma)t} + b_1\mathrm{e}^{(\lambda+\gamma)t} + b_2\mathrm{e}^{(\lambda-\gamma)t},  \label{potential_solution}
\end{equation}
where $\gamma = \sqrt{\lambda^2-\omega^2}$. The last two terms are proportional to the exponentially increasing $\mathrm{e}^{\lambda t}$, so they have non-physical meaning. Consequently, they could not have a role in the measurable $x(t)$.
After the fit of the initial conditions for the measurable quantities, the position, and the velocity, $x(0)=x_0$ and $\dot x(0) = v_0$, keeping the physical solutions, the relevant potential is
\begin{equation}
q(t) = \frac{(\gamma-\lambda)x_0-v_0}{8\gamma\lambda(\lambda+\gamma)} \mathrm{e}^{-(\lambda+\gamma)t} + \frac{(\gamma+\lambda)x_0+v_0}{8\gamma\lambda(\lambda-\gamma)} \mathrm{e}^{-(\lambda-\gamma)t}.  \label{potential_solution_final}
\end{equation}
Now, we are ready to substitute this generator's potential function in the expression of Hamiltonian in Eq. (\ref{oscillator_hamiltonian}). The mathematical calculation results that 
\begin{equation}
H = 0.
\end{equation}
This strange result means that this zero value Hamiltonian is the conserved quantity of the damped oscillator. We may say there is no contradiction in the theory. However, Hamiltonian lost the "total energy of the system" meaning. Despite this situation, we consider the Hamiltonian as an energy-like quantity. We will see that this zero value Hamiltonian enables us to elaborate on the quantization formulation of the dissipative oscillator.

\section{The quantization procedure}  \label{quantization}

To achieve the state equation of the quantized damped oscillator, we need to identify the canonical momenta in the Hamiltonian, $H'$, in Eq. (\ref{oscillator_hamiltonian_by_new_momenta}). In the case of the canonical pair $(P_2,Q_2)$, the physical meaning can be read out from the definition of the potential $q$ in Eq. (\ref{potencial_definition}) and together with the Eqs. (\ref{P_2}) and (\ref{Q_2}) 
\begin{equation}
P_2 \left( = mv = m\frac{\text{d}x}{{\mathrm{d}}t} = \right) = \hbar k , \,\,\,\,\,\,\,\,\,\, Q_2 = x .  \label{P_2_Q_2}
\end{equation}
The momentum $P_2$, and the coordinate $Q_2$ are the usual canonically conjugated pairs. The construction of the momentum $P_1$ and the coordinate $Q_1$ are based on the Eqs. (\ref{P_1}) and (\ref{Q_1}) and a comparison with the momentum $P_2$ and the coordinate $Q_2$ in Eq. (\ref{P_2_Q_2}). The appearing time factor in Eqs. (\ref{P_1}) and (\ref{Q_1}) can be associated with Fourier transformed pairs, i.e.,   
\begin{equation}
P_1 = -\mathrm{i} \hbar k \omega, \,\,\,\,\,\,\,\,\,\, Q_1 = \frac{x}{\mathrm{i}\omega} .  \label{P_1_Q_1}
\end{equation}
\begin{align}
\frac{\partial}{\partial x} = \mathrm{i}k, \,\,\,\,\,\,\,\,\,\, \frac{\partial}{\partial t} = -\mathrm{i}\omega, \nonumber \\ \frac{1}{-\mathrm{i}\omega} = \int ... \,\, {\mathrm{d}}t, \,\,\,\,\,\,\,\,\,\, kx \longrightarrow \frac{1}{\mathrm{i}}.  \label{Fourier_terms}
\end{align}
The terms of the Hamiltonian can be expressed in the operator formulation applying the above rules. The calculation of $P_2$ and the $P_2^2$ go as is usual. We take $P_2$ from Eq. (\ref{P_2_Q_2}) and after that the first Fourier transformed in Eq. (\ref{Fourier_terms}), we obtain
\begin{equation}
P_2 = \hbar k = \frac{\hbar}{\mathrm{i}}\frac{\partial}{\partial x} \,\,\,\,\, \longrightarrow \,\,\,\,\, P_2^2 = - {\hbar^2}\frac{\partial^2}{\partial x^2} .
\end{equation}
The second term includes the $P_2 Q_1$ product. Now, we consider $P_2$ from Eq. (\ref{P_2_Q_2}) and $Q_1$ from Eq. (\ref{P_1_Q_1}), then we apply the third Fourier transform in Eq. (\ref{Fourier_terms}). The detailed steps are shown one by one
\begin{eqnarray}
P_2 Q_1 &=& \hbar k \frac{x}{\mathrm{i}\omega} = \underbrace{\frac{1}{\mathrm{i}\omega}}_{-\int ... \,\, \text{d}t} \underbrace{\hbar k}_{mv} x = -\int mv x \,\, {\mathrm{d}}t  \nonumber \\ 
&=& -\int mx \frac{{\mathrm{d}}x}{{\mathrm{d}}t} \,\, {\mathrm{d}}t = -\frac{1}{2} m x^2 ,
\end{eqnarray}
by which we obtain the second term of the Hamiltonian in the operator formalism
\begin{equation}
- \omega^2 P_2 Q_1 = \frac{1}{2} m \omega^2 x^2 .
\end{equation}
We continue with the $P_1 Q_2$ term. We take $P_1$ from Eq. (\ref{P_1_Q_1}) and $Q_2$ from Eq. (\ref{P_2_Q_2}), and we apply the second and fourth Fourier transform, i.e.,
\begin{equation}
P_1 Q_2 = -\mathrm{i} \hbar k \omega x = -\mathrm{i} \hbar \underbrace{k x}_{1/\mathrm{i}} \underbrace{\omega}_{\mathrm{i} \frac{\partial}{\partial t}}  = \frac{\hbar}{\mathrm{i}} \frac{\partial}{\partial t} .
\end{equation}
The last term of the Hamiltonian includes the $P_2 Q_2$ product. The relevant substitutions come from Eq. (\ref{P_2_Q_2}), and we consider the fourth Fourier transform    
\begin{equation}
P_2 Q_2 = \hbar k x = \hbar \underbrace{k x}_{1/\mathrm{i}} = \frac{\hbar}{\mathrm{i}} ,
\end{equation}
by which we write
\begin{equation}
2 \lambda P_2 Q_2 = 2 \lambda \frac{\hbar}{\mathrm{i}} .
\end{equation}
Taking the Hamiltonian in Eq. (\ref{oscillator_hamiltonian_by_new_momenta}) and substituting these expressions in it, the quantized state equation of the damped oscillator can be formulated as

\begin{equation}
0 = \underbrace{- \frac{\hbar^2}{2m}\frac{\partial^2 \psi}{\partial x^2} + \frac{1}{2} m \omega^2 x^2 \psi + \frac{\hbar}{\mathrm{i}} \frac{\partial \psi}{\partial t}}_{\text{frictionless quantum oscillator}} + \underbrace{2 \lambda \frac{\hbar}{\mathrm{i}} \psi}_{\text{damping term}}.  \label{dissipative_state_equation}
\end{equation}

Similar to the complex absorbing potential in Eq. (\ref{nagative_imaginary_potential}), a non-Hermitian term appears in the state equation. Its role is the same that term generates dissipation in the motion. However, the deduction of the damped state equation comes from a consequent calculation; the complex part of the potential in Eq. (\ref{nagative_imaginary_potential}) is an {\it ad hoc} assumption. We can divide the equation into the undamped quantum harmonic oscillator and the damping part. The damping term also includes the quantum action factor $\hbar$.

\section{Solution of damping wave equation}  \label{path_integral}

\begin{figure}[h!]
\centerline{
\includegraphics[width=0.7\columnwidth]{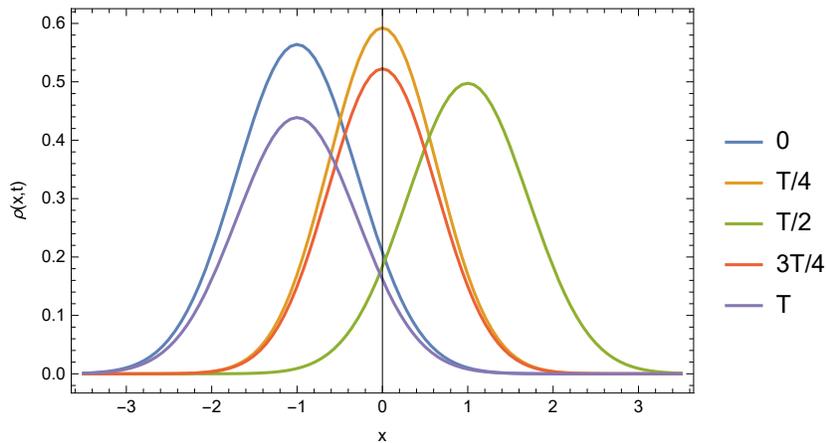}}
\caption{The dissipation caused shape damping of the distribution $\rho(x,t)$ of the damped oscillator in one time period. The applied parameter set is $m = 1$; $\hbar = 1$; $\omega = 1$, $\gamma = 0.8$, $\lambda=0.01$. The time period is $T = {2\pi}/{\omega}$. The peak of the initial distribution is at $x_0 = -1$.}  \label{damped_change_rho}
\end{figure}

The oscillation starts from a normalized Gaussian shape initial wave function, which is the eigenfunction of the lowest-lying energy level of the frictionless case:
\begin{equation}  \label{initial_Gaussian}
\Psi_{0}(x,0) = \sqrt[4]{\frac{ m\omega}{\pi\hbar}} \exp \left( - \frac{m\omega}{2\hbar}(x-x_0)^2 \right)
\end{equation}
with its center position $x_0$. The movement of the undamped oscillator (the frictionless part of Eq. (\ref{dissipative_state_equation})) can be calculated by the Feynman path integral method \cite{feyn1,feyn2,khan,Dittrich2001}. Souriau pointed out a correction that is necessary beyond the first half-period of motion in the integration formula of Feynman \cite{Souriau1975}. However, a further detailed study was to refine the oscillator wave packet motion. Naqvi and Waldenstrøm \cite{Razi2000} introduced a $\gamma \neq 1$ parameter by which the width of the Gaussian wave packet changes periodically in time around the origin (see Fig. 2 in Ref. \cite{markus2016})
\begin{equation}
| \Psi (x,t) |^2 = \frac{1}{\sigma_{x}(t)\sqrt{2}} \times \exp{ \left\{ -\frac{\left[x - x_{0} \cos(\omega t) \right]^2}{2\sigma_{x}^{2}(x,t)} \right\}},  
\end{equation} 
where
\begin{equation}
\sigma_{x}^{2}(x,t) = \frac{\hbar}{2{\gamma}m{\omega}} \left[ \cos^2(\omega t) + {\gamma}^2 \sin^2(\omega t) \right]
\end{equation}
ensures the width-change. \\
The first two terms in Eq. (\ref{dissipative_state_equation}) do not depend on the time, and since the third term includes a first-order time derivative, thus the damping effect can be extracted from the third and fourth terms (see Appendix), i.e.,
\begin{equation}
0 = \frac{\hbar}{\mathrm{i}} \frac{\partial \psi'}{\partial t} + 2 \lambda \frac{\hbar}{\mathrm{i}} \psi' .
\end{equation}
The solution to this equation can be expressed as
\begin{equation}
\psi' \sim \exp \left( -2\lambda t \right).
\end{equation}
Finally, we obtain the exact solution of the quantum dissipative oscillator equation (\ref{dissipative_state_equation}) as
\begin{equation}
| \Psi (x,t) |^2 = \frac{1}{\sigma_{x}(t)\sqrt{2}} \exp{ \left\{ -\frac{\left[x - x_{0} \cos(\omega t) \right]^2}{2\sigma_{x}^{2}(x,t)} \right\}} \times \exp \left( -4\lambda t \right).  
\end{equation}
The time evolution of $| \Psi (x,t) |^2$ is presented in Fig. \ref{damped_change_rho}, as it is similarly experienced from the complex potential approximations in the description of dissipative quantum systems \cite{Razavy2005,markus2016}. However, in the present quantization procedure, the damped oscillator frequency is identical to the frequency of the undamped oscillator. In contrast to the classic case, the damping does not modify the eigenfrequency $\omega$, i.e., damping pertains to the energy and information loss. This fact suggests that the well-known Lorentz distribution in the scattering process does not relate to a frequency shift due to the damping effect on a quantum level. On the other hand, this result is in line with the second quantized solution in Ref. \cite{risken1989}. In this reference, Eq. (A4.21) clearly shows that there is no frequency shift in the case of a damped quantum oscillator but just amplitude damping. The experimental and theoretical motivations for the quantum dissipation can be found in the early Refs. \cite{haken1970,haake1973}. \\ 
A time series of $\rho(x,t)$ with parameters set as $m = 1$, $\hbar = 1$, $\omega = 1$, and  $\lambda=0.01$ is shown in Fig. \ref{damped_change_rho} in which the evolution can be easily followed in one time period.

\section{Conclusion}  \label{conclusion}

We presented a thorough investigation and a solution for the age-old problem, the quantization of the damped oscillator. The wave propagation is dissipative not only on macroscopic but on a microscopic level and nanoscale. The information loss is closely related to the signal distortion, i.e., the energy dissipation. Our method overcomes the limitations of the previous formulations by several means in achieving the damping state equation. We needed to go back to the fundamentals of quantum theory, such as {\it i}) the Lagrangian formulation, {\it ii}) the Hamiltonian canonical description, and {\it iii}) the quantization procedure. Thus, we managed to apply this mathematical framework to the damping oscillator. In possession of the correct Lagrangian description, the way opened toward the quantization procedure. As was shown, a remarkably understandable damping state equation came up. Finally, we calculated the exact solution of this quantum dissipative oscillator equation. We can conclude that a consequent construction of the damping quantum oscillator equation is presented by the exact solution of this damped state equation. However, we emphasize that due to the non-Hermitian complex potential, the probability meaning of the wave function is lost.  The results bring us closer to the understanding of energy dissipation, information loss, and the maximal probability of the recovery of a signal on the microscopic level. Our studies in the area of quantum dissipation require further discussions and examinations.\\

{\bf Appendix} \\

We point out that the solution of the damping quantum oscillator equation can be exactly obtained as a product of the solution of the undamped oscillator and an exponentially decreasing time-dependent function. We start from the dissipative state equation by repeating Eq. (\ref{dissipative_state_equation})
\begin{equation}
0 = - \frac{\hbar^2}{2m}\frac{\partial^2 \psi}{\partial x^2} + \frac{1}{2} m \omega^2 x^2 \psi + \frac{\hbar}{\mathrm{i}} \frac{\partial \psi}{\partial t} + 2 \lambda \frac{\hbar}{\mathrm{i}} \psi.
\end{equation}
Let us denote the solution of the undamped oscillator by  
\begin{equation}
\psi_0(x,t),
\end{equation}
i.e.,
\begin{equation}
0 = - \frac{\hbar^2}{2m}\frac{\partial^2 \psi_0}{\partial x^2} + \frac{1}{2} m \omega^2 x^2 \psi_0 + \frac{\hbar}{\mathrm{i}} \frac{\partial \psi_0}{\partial t}
\end{equation}
is completed. Now, we find the solution to the damped equation in the form 
\begin{equation}
\psi_0(x,t) \exp \left( -2\lambda t \right).
\end{equation}
It is easy to check by substitution that this function is a solution to the problem:
\begin{equation}
0 = \left[ - \frac{\hbar^2}{2m}\frac{\partial^2 \psi_0}{\partial x^2} + \frac{1}{2} m \omega^2 x^2 \psi_0 + \frac{\hbar}{\mathrm{i}} \frac{\partial \psi_0}{\partial t} \right] \times \exp \left( -2\lambda t \right)
+ \frac{\hbar}{\mathrm{i}} \psi_0 \left[ -2\lambda exp \left( -2\lambda t \right) \right] 
+ 2 \lambda \frac{\hbar}{\mathrm{i}} \psi_0 \exp \left( -2\lambda t \right).
\end{equation}
Since the last two terms eliminate each other, we get back the undamped part of the problem. However, we assumed that $\psi_0(x,t)$ is the solution of the undamped motion. Q.E.D. \\

{\bf Author Contributions:} The authors contributed equally to this work. All authors have read and
agreed to the published version of the manuscript. \\  

{\bf Acknowledgment} \\

This research was supported by the National Research, Development and Innovation Office (NKFIH) Grant Nr. K137852 and by the Ministry of Innovation and Technology and the NKFIH within the Quantum Information National Laboratory of Hungary. Project no. TKP2021-NVA-16 has been implemented with the support provided by the Ministry of Innovation and Technology of Hungary from the National Research, Development and Innovation Fund. \\

{\bf Conflicts of Interest:} The authors declare no conflicts of interest.

\end{document}